# DECLINING VOLATILITY, A GENERAL PROPERTY OF DISPARATE SYSTEMS: FROM FOSSILS, TO STOCKS, TO THE STARS

*by* BRUCE S. LIEBERMAN[1] *and* ADRIAN L. MELOTT[2]

[1]Department of Ecology & Evolutionary Biology and Biodiversity Institute; e-mail: blieber@ku.edu

[2]Department of Physics and Astronomy; e-mail: melott@ku.edu

University of Kansas, Lawrence, KS 66045, U.S.A.

**Abstract:** There may be structural principles pertaining to the general behavior of systems that lead to similarities in a variety of different contexts. Classic examples include the descriptive power of fractals, the importance of surface area to volume constraints, the universality of entropy in systems, and mathematical rules of growth and form. Documenting such overarching principles may represent a rejoinder to the Neodarwinian synthesis that emphasizes adaptation and competition. Instead, these principles could indicate the importance of constraint and structure on form and evolution. Here we document a potential example of a phenomenon suggesting congruent behavior of very different systems. We focus on the notion that universally there has been a tendency for more volatile entities to disappear from systems such that the net volatility in these systems tends to decline. We specifically focus on origination and extinction rates in the marine animal fossil record, the performance of stocks in the stock market, and the characters of stars and stellar systems. We consider the evidence that each is experiencing declining volatility, and also consider the broader significance of this.

ONE of the key aspects of research in macroevolution is using the study of evolutionary patterns to extract information about the nature of the evolutionary process (Eldredge and Cracraft 1980; Vrba 1985). Studies of biological patterns in the history of life have led scientists to posit or consider biological rules or laws, e.g. Cope's rule (Stanley 1973; Jablonski 1997), Bergmann's rule (Kurtén 1968), the Zero Force Evolutionary Law (McShea and Brandon 2010), etc. Such laws may or may not be comparable to lawlike behavior documented in the physical sciences, such as Newton's laws, Maxwell's equations, etc. and these topics are discussed in Bromham's (2011) very useful recent review. Sometimes, the quest to identify such general patterns has grasped further, leading to suggestions that there may be structural principles pertaining to the general behavior of systems that lead to similarities in a variety of different contexts. Classic examples especially relevant to biology and paleontology include the well documented descriptive power of fractals (Mandelbrot 1983; Plotnick and Sepkoski 2001), the importance of surface area to volume constraints in organisms (Galileo 1638; Bonner 1988) and the architecture of gothic cathedrals (Gould 1977), the universality of entropy in systems (Brooks and Wiley 1986), and mathematical rules of growth and form in mollusks (Raup 1966), receptaculitids (Gould and Katz 1975), plants (Niklas 1997), and organismal development (Goodman 1994).

Gould (1988*a*) specifically suggested, and McShea and Brandon (2010) amplified the notion, that documenting such overarching principles represented a significant rejoinder to the Neodarwinian synthesis that emphasized the paradigmatic roles of adaptation and competition. Instead, Gould (1988*a*) argued that these principles would speak to the importance of constraint and structure on organismal form and the pathway of evolution. Among Gould's best known example of such a universal principle transcending very disparate systems was his discussion of the congruence between the decline of the .400 hitter in baseball and the geometry of many evolutionary trends: they each exemplified a pattern of declining variance around an unchanged mean (Gould 1988*a*, 1996). This is of course not to suggest that disparate systems always show similar behavior. For instance, it has been demonstrated that cultural evolution is very different from biological evolution (e.g. Eldredge 2011). Still, when commonalities are uncovered it can be relevant to scientists in different fields.

Here we document another potential example of a phenomenon suggesting congruent behavior of very different systems. In particular, universally there has been a tendency for more volatile entities to disappear from systems such that the net volatility in these systems tends to decline. We specifically focus on origination and extinction rates in the marine animal fossil record, the performance of stocks in the stock market, and the character of stellar systems. In these systems high volatility respectively comprises relatively high rates of origination (and extinction) compared to other taxa (Gilinsky 1994), a greater tendency for price to move up (or down) relative to the rest of the stock market, and the star formation (and explosion) rate. We consider the evidence that each is experiencing declining volatility, and also consider the broader significance of this. In an evolutionary context volatility is related to variability, but it is not exactly the same: for instance, an individual species might be less variable, but more likely to speciate (or go extinct) and therefore more volatile. Thus, a volatile entity lacks stability and persistence through time, with volatility often subsuming an aspect of turnover. A perspective focused on volatility helps to explain the evolution of each of the systems we consider because each of these systems is comprised of historical entities.

**DECLINING VOLATILITY IN ORIGINATION AND EXTINCTION RATES**

Declining volatility has been amply documented in the marine fossil record, even as overall diversity has dramatically increased. For instance, Raup and Sepkoski (1982) were the first to show in detail and quantitatively that the background extinction rate declined throughout the Phanerozoic. Subsequent studies have reiterated these results, and shown that origination rates are also declining throughout the Phaernozoic (e.g. Gilinsky 1994; Foote 2000; Miller and Foote 2003; Jablonski 2007; Bambach 2006; Lieberman and Melott 2007). Given that as Eldredge (1979), Stanley (1979), Vrba (1980), Jablonski (1986), Benton (1995) and others have suggested, speciation rates are strongly correlated with extinction rates, it seems quite reasonable that origination and extinction should show similar patterns (although at the large scale origination and extinction rates do not always vary in unison and do not respond synchronously: Kirchner and Weil 2000; Melott and Bamach 2011b; Krug and Jablonski 2012). It is worth mentioning that origination data bearing on this issue do not come from patterns at the species level. Instead, they are derived from genus and family level data such that it is origination rates rather than speciation rates that are being considered. Further, certainly there is debate about the role that the quality of the fossil record may play in artefactually creating this pattern (e.g. Peters and Foote 2002; McGowan and Smith 2008; Lloyd *et al.* 2011), and there are debates about the taxonomic data being used to adduce these patterns which could artefactually bias perceived patterns of origination and extinction (e.g. Patterson and Smith 1987; Adrain *et al.* 1998). Each of these debates is beyond the scope of the present contribution, but on the whole these and other biodiversity patterns preserved in the fossil record seem real (Bambach 2006; Melott and Bambach 2011*a*, 2011*b*; Melott *et al.* 2012).

Gilinsky (1994) was the first to specifically consider in detail (though see also Raup and Boyajian 1988) the phenomenon of declining origination and extinction rates in the context of declining evolutionary volatility through time. His recovered pattern still seems resilient even with refinements to data and timescale: for instance, when considering the fractional rate of origination and extinction of genera derived from Bambach's (2006) version of the Sepkoski dataset (Sepkoski 2002) (Fig. 1) also used in Melott *et al.* (2012). Although the Cambrian values play an important role in the perceived decline in volatility in Figure 1, even when these are excluded a best-fit straight line from the beginning of the Ordovician to the Holocene still shows volatility declining from 0.6 to 0.25. The reason that volatility declines is the long-term evolutionary risk of high volatility. High origination rates are correlated with high extinction rates (Eldredge 1979; Stanley 1979; Vrba 1980), and high volatility increases the chance that the diversity of a taxon will fall to zero, which represents summary extinction, the evolutionary point of no return (Gilinsky 1994). Since high volatility correlates with short geological duration, it is only groups with low rates of origination, and thus low volatility, which persist over long intervals of geological time (Gilinsky 1994). As Gilinsky and Good (1991) pointed out, the evolutionary attributes of low volatility groups like annelids make them most apt to be evolutionary survivors. The negative effects of high volatility seem to be particularly exacerbated by mass extinctions when high volatility groups such as trilobites and ammonites seem to be hit particularly hard (Lieberman and Karim 2010).

Part of the explanation for the pattern of declining volatility over the Phanerozoic resides in the fact that high volatility groups are more likely to be eliminated through time. For instance, the fraction of long-lived genera that persist more than 45 million years has strongly increased (Melott and Bambach 2011*a*). This results in a winnowing out of high volatility groups with seemingly survival of the blandest ensuing. The net effect is that, at least by the measure of survivability and representation in the biota through time, high volatility groups perform poorly relative to their low volatility brethren. Another important aspect of the pattern, however, is that such groups are clearly not being replaced by new high volatility taxa (Gilinsky 1994). This suggests an evolutionary asymmetry whereby at the species-level perhaps high volatility species give rise to both high and low volatility species whereas low volatility species only give rise to low volatility species, resembling the patterns described in Lieberman *et al.* (1993), Lieberman and Vrba (1995), and Gould (2002). Thus, while Figure 1 shows the general overall decline in volatility across the Phanerozoic, this pattern partially (though not solely) results from the disappearance, without replacement, of individual taxa with high volatility.

Ultimately there could be several factors and attributes that make species have high origination and extinction rates and thus be highly volatile, for instance, one is likely small geographic range (Jablonski 1986). Similarly, there are many physical factors that determine the sum obtained from the roll of a pair of dice. The upshot is that in either case the mathematical descriptor provides a good means of characterizing the pattern and also predicting future results (see also Wilkinson 2011).

**DECLINING VOLATILITY AND MAXIMIZING RETURNS FROM THE STOCK MARKET**

The volatility of the price of a stock is often characterized by its ß, which is an implicit measure of volatility that considers how the stock moves relative to the rest of the market. Other measures of volatility do exist, but these are closely correlated with ß (Baker *et al.* 2011). Stocks with a ß > 1 have more volatility and are thus considered more risky. For a long time it was argued based on the Capital Asset Pricing Model (CAPM) that high ß stocks should have higher rates of return than low ß stocks, and CAPM governed the archetypical investment strategies. More recent analyses have belied the validity of the CAPM model (e.g. Karaceski 2002; Ang *et al.* 2006;

Baker *et al.* 2011) and instead show that over the long interval 1968-2008 (from a financial perspective) it is low ß, low volatility stocks that have the highest average returns, and the smallest drawdowns. The difference in performance is actually striking (Fig. 2), with $1US invested in low volatility, low ß stocks in 1968 yielding $10.28 in 2008 while $1US invested in high volatility, high ß stocks yielded just $0.64 (all figures adjusted for inflation, from Baker *et al.* 2011). In an important respect we can think of performance in the stock market as partly analogous to a measure of evolutionary success in the history of life, as by the measure of financial productivity low ß is a much more successful strategy for investors. Possible reasons for the difference in relative performance of low and high ß stocks are discussed in Karaceski (2002) and Baker *et al.* (2011).

The difference in relative performance of low and high ß stocks is particularly exaggerated during severe market downturns and financial crises. For example, during the dramatic bear markets of 1973–1974, 1987, 2000–2002, and 2008 high volatility stocks fared extremely poorly (Baker *et al.* 2011). In a sense, these severe market downturns seem directly analogous to mass extinctions in the fossil record. High volatility entities, be they taxa or stocks, are at extreme risk when the probability of extinction or price decline shoots up, as the taxa already have high probabilities of extinction and the stocks already have high probabilities of price decline. The congruence of these patterns from these two disparate areas is striking.

This is of course not to suggest that the stock market behaves exactly like the history of life. There are certainly important differences. To consider just a few of these, fossil biodiversity closely fits the pattern of a random walk throughout much of the Phanerozoic (Cornette and Lieberman 2004). By contrast, with the stock market, random walks do not prevail and there is some memory to the system: climbing markets tend to climb even higher than they should and also tend to fall even more dramatically during falling markets than they should if they were undergoing a random walk (Lo and Mackinlay 1988; Moskowitz *et al.* 2012). Further, stocks are also not directly analogous to biological entities like species because they do not speciate, except perhaps when a stock 'splits,' although they do go extinct when a company goes bankrupt (except in the very rare instances when that company is deemed 'too big to fail' and receives a bailout). Another important difference pertaining to volatility comes in because new high and low volatility stocks are being created *de novo* all the time and do not typically arise lineally from ancestors. This means that there cannot be an 'evolutionary asymmetry' in the production of stocks. In addition, with stocks there is less of a winnowing out and survival of the blandest effect as high volatility stocks tend to linger around because certain investors behave irrationally and are willing to gamble to secure short-term large gains; in fact, investment firms may be mandated to tacitly adhere to a phenomenon that selects high ß stocks (Baker *et al.* 2011). These differences actually suggest it is even more striking that stock prices display a similar pattern to origination and extinction rates: when it comes to volatility, in both the stock market and the history of life, highly volatile entities perform poorly.

**DECLINING VOLATILITY IN STAR SYSTEMS**

There are significant parallels to biodiversity and stocks in the realm of astrophysics. For example, the star formation rate has declined in our galaxy (Just *et al.* 2011) (Fig. 3) and in the universe as a whole (Lily *et al.* 1996; Madau *et al.* 1996; Johnson *et al.* 2012; Cucciati *et al.* 2012) such that their origination has become less volatile. This is predominantly a consequence of the concentration and depletion of gas that would otherwise form the building blocks of other stars. Just as there are phenomena analogous to origination in star systems, there are phenomena analogous to extinction, and they too show a pattern of decreasing volatility. 'Extinction' of stars can happen when explosive, shorter-lived stars literally wink out of the system in dramatic events such as supernovae and gamma ray bursts (GRBs). Once such entities have gone 'extinct' they can never come back, meaning that the volatile entities are being eliminated from the system. Thus, the proportion of short-lived, more volatile stars is decreasing through time (Gehrels *et al.* 2009). Moreover, through explosive events such as supernovae, material is returned to the interstellar medium, which could serve as the building blocks of new stars, but since the number of these catastrophic events is decreasing the star formation rate falls. The star formation rate is also presumed to have fallen as a consequence of a drop in galaxy merger events, which stimulate star formation in the entrained gas (Lily *et al.* 1996; Madau *et al.* 1996).

There is a distinctive winnowing effect in star systems, also seen in the record of origination and extinction in the fossil record, where it is the more quiescent, longer-lived stars that persist. The upshot is that the decline of volatility has left a quieter, blander universe, just like the record of biodiversity. It is worth noting, however, that the diminution of large core-collapse supernovae and GRBs has likely made the universe a safer, more nurturing place for life to evolve, with fewer higher energy events that might eliminate budding life forms on nascent planets: ramifying astrophysical blandness might well have entailed positive consequences for biological evolution in the universe (Lineweaver *et al.* 2004).

As was the case with the stock market, stars do not behave exactly like taxa, and again there are important differences between astrophysical and biological systems. For instance, the origination of new stars is constrained by the limits of available interstellar material while there is not, at least definitively, such a biological constraint on the evolution of new species, although the notion that biological diversity might fit a logistic equation (MacArthur and Wilson 1967) is worth mentioning. Again, to reiterate our point made earlier, the fact that there are differences between stars and taxa suggests that the similar patterns of stellar evolution and the history of life are even more striking.

**DISCUSSION**

Extensive evidence seems to exist to suggest that declining volatility is a widespread phenomenon spanning and uniting the three disparate areas discussed. The broader question of course is what is the significance of this commonality? Clearly, different mechanisms must be invoked, for instance, to explain declining volatility in star systems and origination and extinction rates in the history of life. Indeed, even within any one of these areas several different mechanisms may be involved in producing the general pattern. It may well be the multiplicity of mechanisms in each area that conspires to produce the generally similar patterns. Undoubtedly another reason though for the similar patterns is that in each area, be it stars, stocks, or taxa, we are dealing with historical entities. The more volatile such entities are, the more likely they are to reach a zero point from which there can be no possibility of return. With taxa this point is extinction, with stars once a stock has experienced a GRB or supernova for all intents and purposes that star is extinct, and in stocks below a certain stock price the firm becomes bankrupt and again for all intents and purposes extinct. The predilection for variable entities to reach a zero point is seemingly exaggerated during times of heightened variability, for instance, mass extinctions and stock market crashes, and for stars early in the history of the universe. It is for these reasons that we can potentially speak of it as a universal principle. Of course, the ability to describe this pattern in general mathematical terms is not unique. For instance, a Poisson process can be used to describe any type of discrete event that happens in continuous time, be it individual humans lining up for tellers in a bank or speciation events occurring across a clade (Feller 1968).

The notion that volatility tends to decline is worth discussing in the context of two principles and themes that Stephen Jay Gould repeatedly emphasized in his works. One is the aforementioned notion that many apparent evolutionary trends, as well as the decline of the .400 hitter in baseball, can be explained as a pattern where variance declines while the mean stays constant (Gould 1988*a*, 1990, 1996, 2002). Although there is a general relationship between this universal principle of Gould and the principle articulated herein, declining volatility is somewhat different because in the case of origination and extinction rates in the history of life not only do these seem to change less through time, but the rates themselves are also declining (Fig. 1). The volatility pattern is more akin to the notion that a landscape might start out topographically complex and due to the forces of erosion that landscape becomes whittled and winnowed away to a flat surface. (This is notably a theme discussed in a different context in Gould 1988b.) In the case of the universe, as it expands the concentration of matter and energy also becomes more diffuse, unless of course the universe were to one day start to shrink and eventually collapse in upon itself, a trajectory which current understanding suggests now seems to be unlikely (Cervantes-Cota and Smoot 2011). In the stock market the metaphor holds in a different way, to the extent that if one is seeking high yields in the market and the height of different parts of the landscape represents the amount of price volatility of individual stocks then one should confine one's investments to stocks in the flatter part of the landscape in order to maximize returns.

Another principle repeatedly articulated by Gould worth discussing vis à vis volatility was the notion that the history of life, specifically in the context of the Cambrian radiation, was marked by initial experimentation with subsequent pruning, retrenchment, and canalization (Gould 1989, 1991, 2002). Certainly the nature of the Cambrian radiation and the validity of Gould's perceived pattern continues to be debated (see Lieberman 2003 and Briggs and Fortey 2005 for reviews), but again, in a broad sense, this principle resembles the principle of declining volatility. However, one area where it differs is that Gould was focusing on the interactions of myriad parts of the organismal developmental system. As these parts became more tightly entwined they might make organismal variability more intrinsically difficult, leading to greater evolutionary constraint. Kaufman (1993) considered this issue of constraint in relation to adaptive peaks, and the higher up an organism sits on an adaptive peak, and the farther apart these peaks are, the harder it would be to move to a different adaptive peak and seize on a new lifestyle. Thus, through time Kaufman (1993) predicted organisms would be locked into particular lifestyles, thereby preventing them from changing. Declining volatility appears different from Gould and Kaufman's proposed biological principle of increasing constraint because it is not about interacting parts becoming more tightly correlated but instead it is more of a statistical principle: greater volatility makes an entity more likely to disappear, and once an historical entity disappears it is gone forever.

Whether the principle of declining volatility can be extended to other systems and thematic areas remains to be seen, and its broader meaning in the context of the three areas mentioned herein still needs to be explored in greater detail. Ultimately, one of the reasons the Neodarwinian synthesis remained incomplete was that it largely restricted its purview to one disciplinary area within the biological sciences, population genetics. Therefore, undoubtedly one way of extending the synthesis and making it more complete is by adding contributions from other biological disciplines including palaeontology, developmental biology, etc. Building on that notion, and following in a long tradition (e.g. Gould 1977, 1988*a*, 1996; Brooks and Wiley 1986; Bonner 1988; Goodman 1994; McShea and Brandon 2010), we firmly believe that identifying general patterns across various areas of scientific and intellectual enquiry is another means of helping to bring the evolutionary synthesis to fruition for 'the wheel of the world swings through the same phases again and again' (R. Kipling) and problems solved by one discipline may represent opportunities for new insights in other areas.

*Acknowledgements*. We thank M. Baker (Harvard Business School), T. Chi (KU School of Business), G. Rudnick (KU), and E. Wiley (KU) for discussions, B. Bradley (Acadian Asset Management), M. Baker, and M. Pettini (U. of Cambridge) for assistance with figures, D. Brooks (U. of Toronto), A. Stigall (Ohio U.), and S. Thomas for comments on an earlier version of this paper, and C. Congreve (KU), C. Myers (KU), and E. Saupe (KU) for inviting us to participate in this special issue. This research was supported by NSF DEB-0716162.

**FIGURE CAPTIONS**

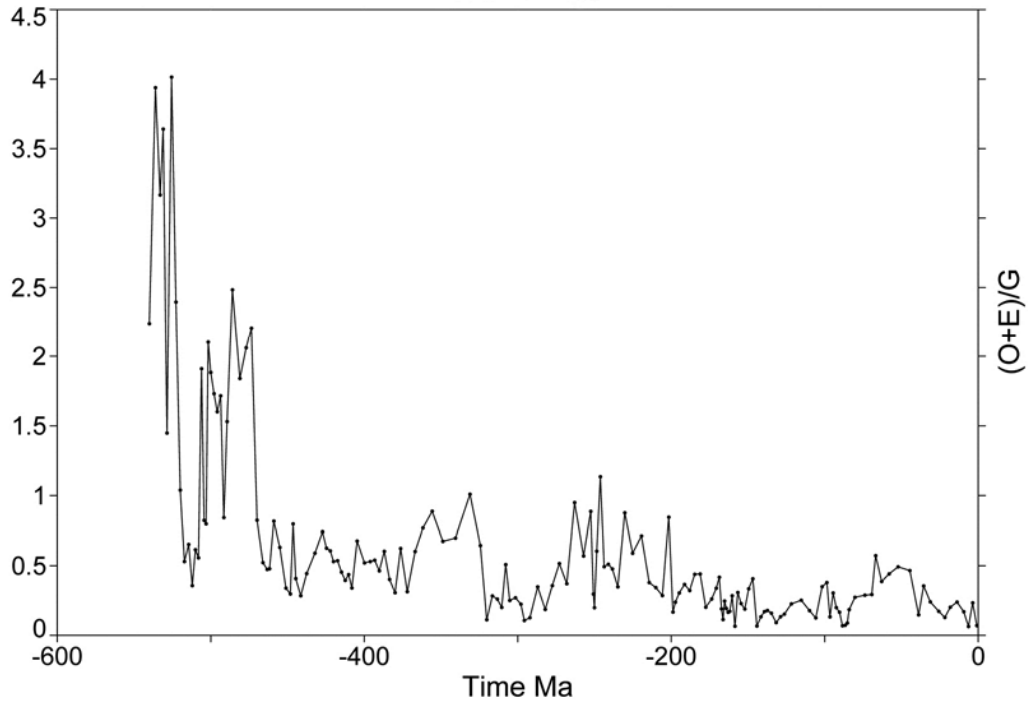

**FIG. 1.** The fractional rate of origination (O) and extinction (E) of genera through time, the total of number of originations plus extinctions divided by the total diversity, derived from Bambach's (2006) version of the Sepkoski dataset (Sepkoski 2002), showing the pattern of declining volatility. Note that O + E is not the usual change in total biodiversity count, which is O – E. Our volatility measure includes one genus being replaced by another. Even without the Cambrian values, volatility still declines precipitously. A similar pattern is found when extinctions are subtracted from originations and divided by total diversity.

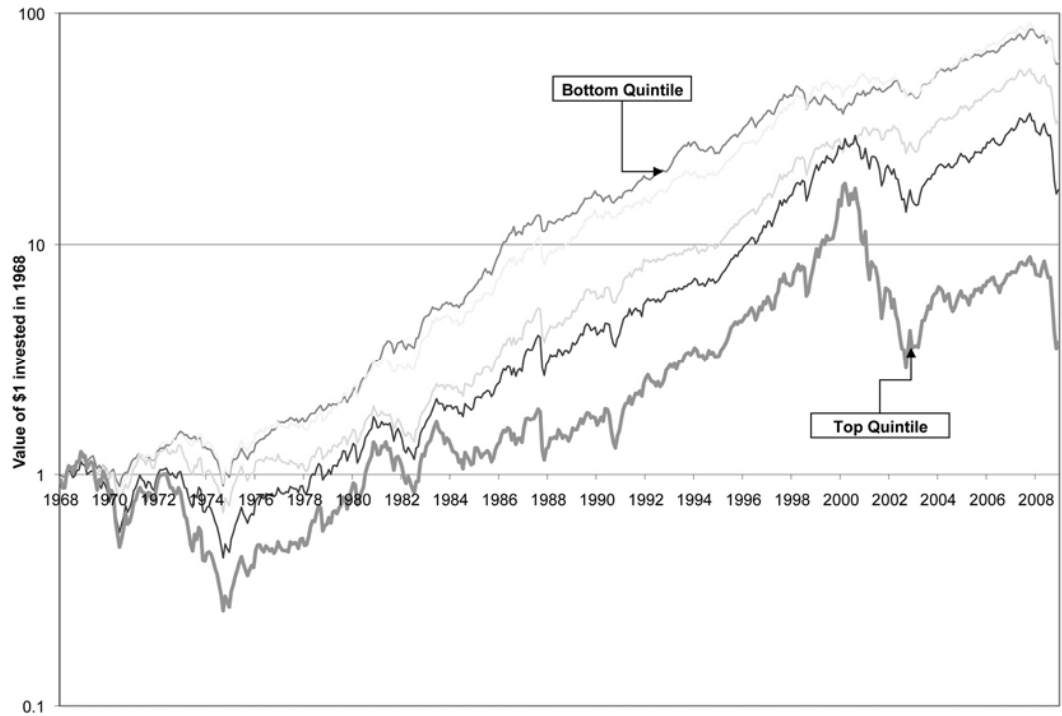

**FIG. 2.** Returns by ß quintiles for all publicly traded stocks (with at least 24 months of return history) from 1968–2008 with $1 invested in January 1968 (not adjusted for inflation). Stocks with the lowest ß and volatility (in the bottom quintile) yield significantly more over time than stocks with the highest ß and volatility (in the top quintile). Copyright 2011, CFA institute. Reproduced from Baker *et al.* (2011) with permission from CFA Institute. All rights reserved.

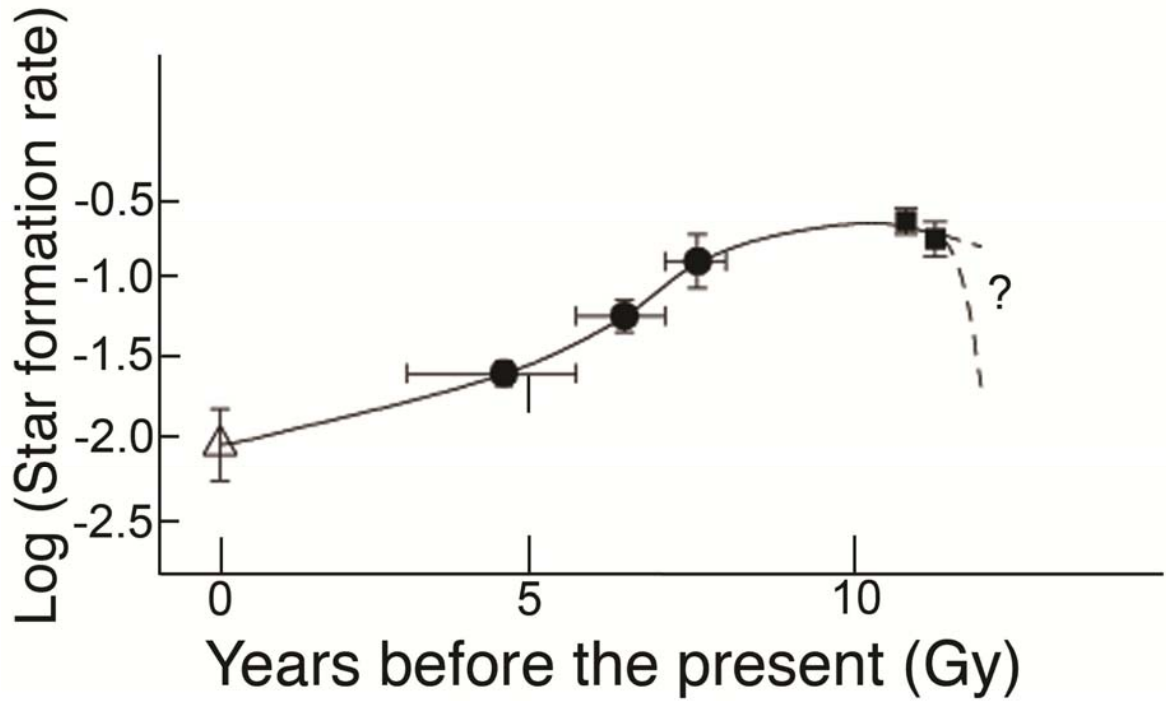

**FIG. 3.** The declining star formation rate over time. This, in conjunction with the elimination of stars due to explosive events, provides an explanation for the decline in volatility in stellar populations through time. Modified from Pettini (2004), used with permission.